# Fluctuations of a Photon Bose-Einstein Condensate Coupled to a Reservoir: Describing Coherence Properties in a Free-Energy Model

Martin Weitz[1], Andreas Redmann[1], Riccardo Panico[1,*], Leon Kleeblank[1], Kevin J.H. Peters[1], Frank Vewinger[1], and Julian Schmitt[1,2]

[1]*Institut für Angewandte Physik, Universität Bonn, Wegelerstr. 8, 53115 Bonn, Germany*

[2]*Kirchhoff-Institut für Physik, Universität Heidelberg, Im Neuenheimer Feld 225a, 69120 Heidelberg, Germany*

Photons are mutually nearly noninteracting particles, so thermalized photon ensembles are commonly obtained not from direct particle-particle-interactions but rather from contact with matter, which can constitute a reservoir for the photon gas. We develop a theory model for photons in a material-filled (e.g. liquid dye) optical microcavity, with the aim to study the fluctuation properties using a free-energy description for noninteracting photons coupled to a reservoir of material electronic excitations. To begin with, we use a single mode description for the condensate. For a small relative size of the material reservoir, corresponding to the canonical regime, condensate number fluctuations are small, and the derived free energy landscape takes the usual Mexican-hat shaped form such that spontaneous symmetry breaking occurs. In contrast, for a large relative size of the reservoir, corresponding to the grand canonical regime, fluctuations become as large as the average particle number. We show that the resulting free energy landscape acquires a bowl-shaped form, with a single minimum at the origin. Thus, a macroscopic occupation of the ground state (i.e., Bose-Einstein condensation) in the absence of spontaneous symmetry breaking is expected. We also provide a model for the treatment of a photon gas trapped in a box-shaped potential with spatially distributed coupling to a reservoir. The model predicts, for example, a statistically fluctuating pattern of islands with long-range coherence, resembling transient microcondensates.

## I.) Introduction

Bose-Einstein condensation, the phase transition of integer spin (bosonic) particles to a macroscopically occupied ground state, has to date been realized in a wide range of physical systems, from cold atoms over exciton-polaritons to photons in material-filled optical microcavities [1-5]. On the theoretical side, Bose-Einstein condensation has been proposed by Einstein as a transition to a state with drastically enhanced single-particle occupation [6].

Subsequent theoretical work, notably by Bogoliubov, Penrose, Onsager, and Anderson, has then formed the picture of spontaneous symmetry breaking accompanying Bose-Einstein condensation [7-10]. Experimentally, the emergence of macroscopic phase coherence, as a consequence of spontaneous symmetry breaking that leads to the condensate choosing a global phase, has been experimentally observed in several physical systems [10-13].

Bose-Einstein condensation of photons has been realized e.g. in dye-solution filled optical microcavities [5, 14-16]. Beyond dye microcavity systems, Bose-Einstein condensation of photons has also been observed in optical fibers [17], and more recently in semiconductor microcavities using VCSEL-like experimental configurations [18,19]. Other than in the case of blackbody radiation, in such two-dimensional photon gas systems, with thermalization being achieved by absorption re-emission processes on the dye molecules, the chemical potential of the photon gas becomes freely tuneable, and a thermodynamic phase transition to condensation can be observed. The cavity mirrors, with a mirror distance in the micrometer regime, due to their small spacing impose an upper limit to the wavelength that fits in the cavity, corresponding to the introduction of a low-frequency cutoff. The longitudinal optical wavevector is frozen out, which makes the system two-dimensional, and the dispersion becomes quadratic, i.e. particle-like.

Thermalization by absorption and re-emission processes on the dye molecules implies that the photo-excitable molecules do not only act as a temperature reservoir for the photon gas, but also as a particle reservoir, a situation that for a large reservoir is described by the grand canonical ensemble. In this physical situation, unusually large particle number fluctuations occur in the condensed state [20-26], as have been experimentally observed in the dye-microcavity system [27]. The magnitude of the fluctuations can be tuned by varying the relative size of the condensate and the dye reservoir. In so far carried out experimental work, despite relatively large statistical number fluctuations, which were measured by recording the second-order correlations, the condensate was observed to still exhibit a relatively large degree of first-order coherence, without experimentally resolving phase fluctuations in the deep grand canonical statistical regime [13]. In earlier theory work, an effective non-equilibrium free energy model has been used to describe formation jitter in a dye microcavity photon condensate [28]. Further, several theory models have been developed to describe the number statistics of photon condensates coupled to a reservoir of dye molecules, using e.g. master equation, super-statistical, and quantum trajectory approaches [23, 24, 29, 30]. In other work [31], it has been argued that grand canonical fluctuations of condensates are not possible, based on the assumption that spontaneous symmetry breaking would be a necessary condition for condensation. However, in [26] it has been pointed out that this conclusion relies on the Bogoliubov quasi-average construction [7], which may not be applicable for such fluctuating condensates.

In the present work, we develop a free energy model for photons coupled to the reservoir of photo-excitable material excitations, e.g. dye molecules, using a single mode description for the condensate. To account for reservoirs of variable size, we employ a model in which both the condensate photons and the dye molecules are treated as constituents of the system within an "extended" canonical ensemble description. The condensate photons interact with the dye via repeated absorption and re-emission processes, establishing exchange between photons and dye excitations. Through the thermal contact of the dye molecules with the surrounding solvent, the combined system - comprising photons and dye excitations - is coupled to a heat bath.

By varying the relative numbers of dye molecules and photons, we can continuously interpolate between two limiting cases, expressed here in the more common terminology wherein only condensate photons are treated as the system. In this way, we smoothly tune from a regime where the condensate photons are effectively coupled to an almost canonical dye reservoir, and photon number fluctuations are suppressed, to the grand-canonical regime, where fluctuations approach the mean photon number. The latter situation is fulfilled when the effective reservoir size $M_{eff}$ is more than quadratically large compared to the condensate population. The normalized zero-delay intensity correlation smoothly varies from $g^{(2)}(0) = 1$ (canonical) and $g^{(2)}(0) = 2$ (grand-canonical), as is well known from earlier works [23, 24, 29, 30].

Within our model, we derive the free-energy landscape of the non-interacting photon gas as a function of the complex-valued condensate field. For a small dye reservoir ("canonical regime"), the free energy exhibits the usual Mexican-hat shape, which implies spontaneous symmetry breaking and aligns well with earlier experimental findings [13]. Such a free energy landscape is well known in the description of interacting quantum gases, as well as in the presence of losses, such as in laser physics; see also [28]. On the other hand, for a large relative effective dye reservoir size ("grand canonical regime"), we predict a bowl-shaped free energy landscape with a single minimum at the origin. In this regime, Bose-Einstein condensation in the absence of spontaneous symmetry breaking is expected. We further generalize our model to a spatially dependent version. In this setting, a spatially dependent photon density for which not the whole dye reservoir acts as a reservoir simultaneously is assumed, but rather the degree of interconversion between photons and dye excitations can spatially vary over the two-dimensional resonator plane. As a consequence, e.g. transiently emerging patches with long-range coherence ("microcondensates") are expected, triggered by thermal fluctuations.

In the following, chapter II gives a single-mode version of a free energy model, and chapter III a model with spatially distributed coupling to a reservoir. Finally, chapter IV gives conclusions and an outlook.

## II.) Single-mode free energy model for photons coupled to a variable size reservoir

We use a model that builds on previous work treating fluctuation properties of photons in a dye-filled microcavity, where radiative contact to the photo-excitable dye molecules enables thermalization of the photon gas confined in the microcavity [23], see also Fig. 1a. Photons are treated as noninteracting particles, and we assume the lossless limit; for a discussion of the influence of losses on the fluctuation dynamics see e.g. [32]. The single mode description for the condensate photons coupling by effective particle exchange to the dye excitations that is used in this chapter assumes a photon gas confined in a harmonic trap realized by curved mirrors, as has been used in earlier work to describe the condensate photon statistics [23, 29, 30, 33]. The model further assumes that the confinement is sufficiently tight for the photon oscillation period to be shorter than the reabsorption time by the dye molecules, so that local variations in photon–dye interconversion can be neglected, and that the populations of the excited photon modes remain much smaller than the size of the dye reservoir such that they do not need to be considered when describing the condensate-mode photon statistics. In the following, $M$ denotes the number of molecules interacting with photons in the condensate mode, of which $M_g$ and $M_e=M-M_g$ are in the electronic ground and excited state, respectively, and $N_0$ is the condensate photon number. The interconversion of photons and dye electronic excitations by absorption and emission into this mode can be described as a photochemical reaction, as represented in Fig. 1b, which leaves the excitation number $X := N_0 + M_e$, the sum of the condensate particle number and dye excitations, constant.

We begin by deriving the probability distribution $P(M_e)$ for finding $M_e$ molecules in the electronically excited state, which via the condensate number $N_0 = X-M_e$ directly relates to the photon number distribution. Here, we neglect the polarization degeneracy. The number of configurations of finding a given number of molecules in the excited state $\Omega(M_e)$, when accounting for the different configurations $\{f_1,\ldots f_M\}$ of the molecules, where $f_i=0$ ($f_i=1$) if the i-th molecule is in the ground (excited) state, and $M_e = \sum_{i=0}^{M} f_i$, is $\Omega(M_e) = M!/[(M_e!\,(M-M_e)!]$. From this, we directly obtain the probability distribution $P(M_e) = \Omega(M_e)/Z \cdot \exp[-E(M_e)/k_B T]$, where $Z$ is the partition function, and $E(M_e)$ the energy of having such a number of dye excitations (measured, as in [33], relative to the state with $M_e = X$). Here $E = (X-M_e)\hbar\Delta$, where $\Delta=\omega_c-\omega_{ZPL}$ denotes the detuning of the cavity cutoff frequency $\omega_c$ (i.e. the condensate frequency) to the zero-phonon line $\omega_{ZPL}$ of the dye (for rhodamine 6G dye: $\omega_{ZPL}\simeq 2\pi c/545$ nm). In our model, we apply the canonical ensemble distribution function and consider both photons and dye electronic excitations as the system, while the solvent of the dye solution serves as a heat bath (see Fig. 1c and the discussion in Chapter 1). The obtained probability distribution can be written as

$$P(M_e) = \frac{1}{Z}\exp\left[-\frac{E(M_e)-k_B T\ln\Omega(M_e)}{k_B T}\right]. \quad (1)$$

Using the entropy $S = k_B \ln \Omega(M_e)$ and the free energy $F = E-T\cdot S$, we obtain the desired expression for the probability distribution in terms of the free energy $F(M_e)$:

$$P(M_e) = \frac{1}{Z}\exp\left[-\frac{F(M_e)}{k_B T}\right] \quad (2)$$

This expression can equivalently be written as a function of the condensate particle number, yielding $P(N_0) = 1/Z \cdot \exp[-F(N_0)/(k_B T)]$. The corresponding partition function is given by $Z = \sum_{M_e=0}^{X} \exp[-F(M_e)/(k_B T)] = \sum_{N_0=0}^{X} \exp[-F(N_0)/(k_B T)]$. The expression for the photon number distribution $P(N_0)$ is consistent with results obtained in earlier works [23, 33], such that previously derived results for the photon statistics can be reproduced (see also the representation of Fig. 1d). By replacing the sum over the photon number by an integral in the above formalism, we can express the average photon number as $\langle N_0 \rangle \simeq 1/Z \cdot$ $\int_0^X N_0\ \exp[-F(N_0)/(k_BT)]\,dN_0$ and the variance as $<N_0^2> \simeq 1/Z \cdot \int_0^X N_0^2\ \exp[-F(N_0)/(k_BT)]\,dN_0$, where the partition function is accordingly $Z \simeq \int_0^X \exp[-F(N_0)/(k_BT)]\,dN_0$. Furthermore, when using the Stirling approximation $\ln(x!) \simeq x\cdot\ln(x) - x$ , the free energy can be approximated by

$$F \simeq (X - M_e)\hbar\Delta - k_BT[M\ln(M) - M_e\ln(M_e) - (M - M_e)\ln(M - M_e)]. \qquad (3)$$

Next, we search for a minimum of the free energy, and aim for an expansion around the minimum value. The derivative $dF/dM_e = -\hbar\Delta + k_BT \cdot \ln[(M_e/(M - M_e)]$ vanishes at $M_{e,min} = M/\left[1 + \exp\left(-\frac{\hbar\Delta}{k_B\mathrm{T}}\right)\right]$, which corresponds to a minimum due to $d^2F/dM_e^2 > 0$. Around this minimum, a quadratic expansion yields:

$$\begin{aligned} F(M_e) &\simeq F\left(M_{e,min}\right) + \frac{(M_e - M_{e,min})^2}{2}\cdot\frac{d^2F}{dM_e^2}\bigg|_{M_{e,min}} \\ &= F\left(M_{e,min}\right) + \frac{(M_e - M_{e,min})^2}{2}k_BT\frac{M}{(M - M_{e,min})M_{e,min}}. \end{aligned} \qquad (4)$$

When using this expansion, one should be aware that due to $X = N_0 + M_e$ and only photon numbers $N_0 \geq 0$ being physical, the maximum physically relevant value the electronic upper state population can take is $M_e = X$. We now proceed to consider the two limiting cases of $X > M_{e,min}$ and $X < M_{e,min}$. We hereby note that fluctuations of the free energy will be of order $k_BT$.

For $X > M_{e,min}$, when additionally assuming that we are far enough away from $M_{e,min} = X$ that $F(X) - F(M_{e,min}) >> k_BT$, we have $\overline{M_e} = M_{e,min}$ and the minimum of the free energy with $N_0 = X - M_e$ is clearly at a non-vanishing photon number. We here do not expect the photon probability distribution to reach down to near zero photon number, see also Fig. 2a. The quadratic expansion can be used to determine the variance of the photon number distribution. Using $P(M_e) = 1/Z \cdot \exp[-F(M_e)/(k_BT)] \propto \exp[-(N_0 - \overline{N_0})^2/(2M_{\mathrm{eff}})]$ , where $M_{\mathrm{eff}} = \overline{M_e}(M - \overline{M_e})/M \equiv M/\{2 + 2\cosh[\hbar(\omega_c - \omega_{\mathrm{ZPL}})/(k_BT)]\}$ denotes the effective reservoir size and $\overline{N_0} = X - \overline{M_e}$ the average photon number, we find $<(N_0 - \overline{N_0})^2> = M_{\mathrm{eff}}$. This yields an expected rms photon number fluctuation $\Delta N_0 = \sqrt{M_{\mathrm{eff}}}$ . With $F(X) - F(M_{e,min}) >> k_BT$ one from eq. (4) also finds that $\overline{N_0} >> \sqrt{M_{\mathrm{eff}}}$ , a condition that has in earlier work already been derived for the canonical photon BEC statistical distribution [23], as the here relevant situation. With the above formula $\Delta N_0 = \sqrt{M_{\mathrm{eff}}}$ this means we have $\Delta N_0 \ll \overline{N_0}$, i.e. fluctuations are small, as expected in this regime. When expressing the free energy in terms of a condensate wavefunction φ, with $N_0 = |\varphi|^2$, we obtain in this limit from eq. (4) the usual Mexican-hat type variation:

$$F(\varphi) \simeq F(\varphi_{Fmin}) + \frac{\left[|\varphi_{Fmin}|^2 - |\varphi|^2\right]^2}{2M_{\mathrm{eff}}} k_B T, \quad (5)$$

which exhibits a ring-shaped minimum at a nonzero value of $|\varphi|$, given by $|\varphi_{Fmin}| = \sqrt{X - M_{\mathrm{e,min}}}$ , see Fig. 2d. In this situation, the U(1) symmetry is expected to be spontaneously broken.

Next, we consider the opposite case of $X < M_{e,min}$, meaning that the minimum of the free energy $F(M_e)$ is not reached at the given number of $X$ excitations, and we again assume that this minimum is relatively far apart from $M_{e,min} = X$, with $F(X) - F(M_{e,min}) \gg k_B T$, as the opposite limit of the case considered above. The smallest possible value of the free energy in the accessible parameter space then is at $M_e = X$, corresponding to a photon number $N_0 = 0$ (see also Fig. 2c). In the relevant energy range $F(M_e) - F(X) \simeq k_B T$, the free energy can be linearly expanded around this minimum possible value. Expressed as a function of the condensate photon number $N_0$ this yields

$$F(N_0) \simeq F(0) + N_0 \cdot \frac{(M_{e,min} - X)\, M}{M_{e,min}\,(M - M_{e,min})} k_B T \quad (6)$$

Using this expression, the corresponding photon probability distribution $P(N_0) = 1/Z \cdot \exp[-F(N_0)/(k_B T)]$ allows us to readily find the average photon number and the variance. This yields $\overline{N_0} = M_{e,min}\,(M - M_{e,min})/\left[(M_{e,min} - X)\, M\right]$ and $\Delta N_0 = \sqrt{< N_0^2 > - < N_0 >^2} = \overline{N_0}$, as expected for the grand canonical statistical regime. From $F(X) - F(M_{e,min}) \gg k_B T$ one here also finds $M_{e,min} - X \gg \overline{N_0}$ . Consequently, $\frac{M_{e,min}\,(M - M_{e,min})}{M} \simeq M_{\mathrm{eff}} = \overline{N_0}(M_{e,min} - X) \gg \overline{N_0}^2$, i.e. we recover the usual condition of the requirement of a large effective relative reservoir size $M_{\mathrm{eff}} \gg \overline{N_0}^2$ in this regime. Physically, condensate photons here are solely generated by the statistical fluctuations of the reservoir, i.e., when assuming this very form of the free energy, the system would remain in the zero photon state in the absence of thermal fluctuations. When again expressing the condensate photon number as the modulus square of a condensate wavefunction φ, with $N_0 = |\varphi|^2$, we have:

$$F(\varphi) \simeq F(0) + B \cdot |\varphi|^2 \qquad (7)$$

with $B = \left(M_{e,min} - X\right) M / \left[M_{e,min}\left(M - M_{e,min}\right)\right]$, see also Fig. 2f. The free energy now has a minimum at the origin for $|\varphi|=0$. Hence, from inspection of the free energy landscape we do not expect the condensate to choose a stable macroscopic phase, there is no corresponding argument for spontaneous symmetry breaking to occur upon the transition from a thermal state to such a condensate.

When regarding dynamics and in particular the phase evolution of the condensate, one should however be aware that this is not necessarily determined by the free energy landscape alone. On the other hand, numeric simulations [13] in the far grand canonical limit predict large excursions to very low and also vanishing photon numbers, where in the latter case a complete restart of the condensate from spontaneous emission occurs with a newly selected random phase and also in the former case the phase of the condensate becomes highly susceptible to such noise. Physically, the condensate in this regime is triggered from thermal fluctuations of the reservoir, which goes along with frequent annihilations and subsequent restarts of the condensate, see also Fig. 2c. In earlier theory work, the timescale on which the first-order coherence function $g^{(1)}(\tau)=\exp(-t/\tau_1)$ temporally decays has in the limit of $M_{\mathrm{eff}} \gg \overline{N_0}^2$ been predicted to be $\tau_1= 2\bar{n}/(B_{\mathrm{em}}\overline{M_e})$, which is only twice the characteristic timescale of intensity fluctuations $\tau_2= \bar{n}/(B_{\mathrm{em}}\overline{M_e})$, with $g^{(2)}(\tau)=1+\exp(-t/\tau_2)$ [30, 34]. The usual effect of phase stabilization for a condensate, resulting that $\tau_1 \gg 2\tau_2$, here is not at work. The coherence properties of a grand canonical photon condensate resemble those of a thermal source, where one also has $\tau_1= 2\tau_2$. We point out that this is fully consistent with the picture of Bose-Einstein condensation described above here occurring in the absence of spontaneous symmetry breaking, because the expectation value of the ground state photon number $\overline{N_0}$ nevertheless is macroscopic.

The condensate average photon number is considered to be a good candidate for the order parameter in this phase when searching for a quantity that is more long-lived than the wavefunction. It shows the usual dependence of transitioning to macroscopic occupation when cooling (or increasing the total particle number correspondingly) through the Bose–Einstein condensation phase transition at the critical temperature $T_c$. The latter is given by $T_c = \hbar\Omega\sqrt{6\overline{N}}/(\pi k_B)$, where $\Omega$ denotes the harmonic trapping frequency induced by the mirror curvature, and $\overline{N}$ is the total photon number. We note that a model distinguishing between coherent and incoherent fractions of the populations respectively, has recently been discussed in the context of photon condensates [35].

We next discuss where to expect the transition between the two described regimes of a Mexican-hat and a single-parabola form of the free energy landscape. This transition occurs when $M_{e,min} = X$ (see Fig. 2b), such that the minimum of the parabolic free energy curve $F(M_e)$ is at $N_0 = X - M_e = 0$, i.e. vanishing condensate photon number. A quadratic expansion of the free energy around $M_e = X$ (i.e., as in Eq. (4) with $M_{e,min} = X$) then yields the following expression, expressed in terms of the condensate photon number $N_0$:

$$F(N_0) \simeq F(0) + \frac{N_0^2}{2M_{\text{eff}}} k_B T. \quad (8)$$

When expressing the free energy as a function of the wavefunction φ we obtain $F(\varphi) \simeq F(0) + |\varphi|^4 \cdot k_B T/2M_{\text{eff}}$, i.e. a quartic dependence. This is characteristic of the transition between Mexican-hat and a bowl-shaped free-energy landscapes, where the curvature at the origin vanishes and the free-energy landscape becomes locally flat (see Fig. 2e). For the variance and the average photon number at $M_e = X$ we find $\langle N_0^2 \rangle = M_{\text{eff}}$ and $\overline{N_0} = \sqrt{2/\pi} \cdot \sqrt{M_{\text{eff}}}$, which yields $\Delta N_0 = \overline{N_0}\sqrt{(\pi - 2)/2}$. The corresponding zero-delay second-order coherence function is $g^{(2)}(0) = \langle N_0^2 \rangle / \overline{N_0}^2 = \pi/2 \simeq 1.57$, a value that has in earlier work also been identified as the point at which the photon statistics changes from being Poissonian to Bose-Einstein-like for the canonical and grand canonical regimes, respectively [23]. The variation of the free energy landscape around the critical point at $M_{e,min} = X$ assumes the typical form as for a continuous phase transition in the thermodynamic limit, for which $\overline{N_0}^2/M_{\text{eff}}$ should be kept fixed to leave the shape of the curve as well as the fluctuation level, i.e. $g^{(2)}(0)$, constant [23, 36].

It is instructive to examine the relevant physics as a function of temperature $T$, with the molecule number $M$ and the total photon number $\overline{N}$ kept constant, following Ref. [23]. As the temperature is lowered below the critical temperature $T_c = \sqrt{6\overline{N}} \cdot \frac{\hbar\Omega}{\pi k_B}$ for Bose-Einstein condensation in the two-dimensional harmonically trapped system, the condensate mode population $\overline{N_0}$ grows following $\overline{N_0}(T) = \overline{N} - \frac{\pi^2}{6}\left(\frac{k_B T}{\hbar\Omega}\right)^2$. For the typically used negative dye-cavity detunings ($\Delta = \omega_c - \omega_{\mathrm{ZPL}} < 0$), in addition the effective reservoir size $M_{\mathrm{eff}}$ decreases as the excited state population reduces. Below a temperature $T_x$ , defined by $\bar{N} - \frac{\pi^2}{6}\left(\frac{k_B T_x}{\hbar\Omega}\right)^2 = \sqrt{2/\pi}\cdot\sqrt{M_{\mathrm{eff}}(T_x)}$ , where the left-hand side of the equation is equal to $\overline{N_0(T_x)}$, the condensate becomes so large that the grand canonical limit is no more fulfilled[1]. As the free energy landscape acquires a Mexican-hat shaped form we enter the "canonical" BEC phase, with macroscopic phase coherence emerging and the condensate wavefunction φ becoming a meaningful order parameter. Note that the grand canonical phase can also reach down to very low temperatures, i.e. we can have $T_x \ll T_c$, provided that the molecule number is sufficiently large: $M \gtrsim \pi \bar{N}^2\{1 + \cosh[\hbar\Delta/(k_B T)]\}$, where we have used $\overline{N_0} \simeq \bar{N}$. This condition on the molecule number becomes easier to reach for small dye-cavity detunings.

In the future, it would also be interesting to investigate the influence of residual effective photon interactions. As long as the interaction energy remains below $k_B T$, one expects the grand canonical number statistics still to well apply. Additionally, real experiments will take place in the presence of finite system loss. To model this, we have performed stochastic phase-space simulations of the driven-dissipative dynamics, based on a Wigner representation of the field. In the limit of small loss with respect to thermalization, i.e. when the system is close to thermal equilibrium, the numeric simulations for the free energy yield results fully consistent with the here described analytic model. Further details will be reported in future work. For experimental studies of the loss of macroscopic coherence expected in the grand canonical regime, the characteristic timescales of the first and the second-order coherence times should be determined, and extended deeper into the grand canonical limit than in so far carried out related experimental work [13].

## III.) Model for a spatially distributed coupling to a reservoir

We next consider the case of a dye microcavity photon gas where a spatially variable degree of interconversion between photons and dye electronic excitation becomes relevant. In other words, we allow for a degree of interconversion that depends on the position within the resonator plane (Fig. 3a). In the following we extend our previous model to also account for the spatial degrees of freedom. Assume that the two-dimensional photon gas in the microresonator is confined by a hard-wall box-potential, of length $L$ in both the $x$ and $y$ transverse dimensions of the cavity plane. For related experimental work studying two-dimensional box-trapped photon gases, see e.g. [37,38].

---

[1] We note that the analytic estimate of $T_x$ given in [23] yielded a relation between $\overline{N_0(T_x)}$ and $\sqrt{M_{\mathrm{eff}}(T_x)}$ that differs from our result by a factor $\sqrt{2/\pi}$ . However, our result agrees well with the more accurate numerical calculations reported in the same reference.

For such two-dimensional noninteracting homogeneous systems, other than in the harmonically trapped case, long range phase fluctuations occur in the quantum degenerate regime. In this regime, i.e. for $n\lambda_{th}^2 \gtrsim 1$, where $n$ is the area density and $\lambda_{th}=h/\sqrt{2\pi m k_B T}$ the thermal wavelength, when evaluating the first-order correlation function $g^{(1)}(\vec{r})$ from the Fourier-transform of the momentum-space picture density distribution, as described in [39], one finds in the limit $r >> \lambda_{th}$ that $g^{(1)}(r) \propto \frac{1}{\sqrt{r}} \exp(-r/\xi)$, where $\xi$ denotes the correlation length. In the above formula for the thermal wavelength of the two-dimensional optical quantum gas, $m=h/(c\cdot\lambda_{cutoff})$ is an effective photon mass, where $\lambda_{cutoff}$ denotes the low-frequency cutoff wavelength of the microcavity and $c$ is the speed of light in the medium. Consider now a gas with average density $\bar{n} \lesssim 1/\lambda_{th}^2$. If a grand canonical fluctuation locally increases the density around $\vec{r} = \vec{r_0}$ such that here quantum degeneracy, with $n\lambda_{th}^2 \gtrsim 1$, is reached, a localized wavepacket with long-range (beyond $\lambda_h$) coherence develops. Phase coherence then is established over a spatial extent set by the correlation length $\xi$, and the localized wavepacket behaves as a (transient) microcondensate. A useful ansatz for the corresponding wavefunction is: $\varphi(\vec{r}) \propto \frac{1}{\sqrt{\vec{r}-\vec{r_0}}} \exp[-(\vec{r}-\vec{r_0})/\xi]$ Note that the ansatz is motivated by the decay of the correlations in an equilibrated quantum-degenerate two-dimensional Bose gas in a homogeneous system, with a correlation length that depends on the chemical potential.

Let us next evaluate the free energy associated with a local interconversion of the grand canonical system from dye electronic excitations to photons, resulting in such a local microcondensate over a patch size of area $\sim \xi^2$. In the following, $\rho_g(\vec{r})$ and $\rho_e(\vec{r})$ denote the local area densities of ground and excited state molecules, with $\rho_g(\vec{r}) + \rho_e(\vec{r}) = \rho = M/L^2$. In our model we assume that photons remain stationary in the two-dimensional resonator plane, which requires sufficiently rapid reabsorption of the dye. We then have $n(\vec{r}) + \rho_e(\vec{r}) = X/L^2$, meaning that the density of excitations is conserved upon (local) interconversion from photons to dye excitations. When in addition accounting for the kinetic energy to yield a Ginzburg-Landau Ansatz for the free energy of the system in an area $\xi^2$ of spatially uniform interconversion, we in analogy to eq. (3) arrive at:

$$F(\rho) \simeq -\frac{\hbar^2}{2m}\int d^2r|\nabla\varphi|^2 - \xi^2\left(\frac{X}{L^2} - \rho_e\right)\hbar\Delta$$
$$-k_BT\left[\xi^2\rho\ln(\xi^2\rho) - \xi^2\rho_e\ln(\xi^2\rho_e) - \xi^2(\rho-\rho_e)\ln(\xi^2(\rho-\rho_e))\right]. \quad (9)$$

The first kinetic term $\sim \hbar^2/(2m\xi^2)$ is suppressed with respect to the second (and the final) term ($\sim k_BT$) by a factor $(\lambda_{th}/\xi)^2$, and can therefore be neglected. For a given $\xi$, the free energy $F(\rho_e)$ exhibits a minimum at $\rho_{e,min} = \rho\frac{1}{1+\exp(-\hbar\Delta/k_BT)}$, and a quadratic expansion around this minimum yields:

$$F(\rho_e) \simeq F(\rho_{e,min}) + \frac{(\rho_e-\rho_{e,min})^2}{2}k_BT\xi^2\frac{\rho}{(\rho-\rho_{e,min})\rho_{e,min}}. \quad (10)$$

Since fluctuations in the free energy are of order $\Delta F \simeq k_BT$ and density fluctuations in the excited state are directly converted to photon gas density fluctuations, we have $\langle(\rho_e - \rho_{e,min})^2\rangle = \Delta n^2$. Assuming that $\rho/\left[(\rho-\rho_{e,min})\rho_{e,min}\right] \simeq \rho/[(\rho-\overline{\rho_e})\overline{\rho_e}] =: 1/\rho_{eff}$, we find $\Delta n = \sqrt{2\rho_{eff}}/\xi$, where $\rho_{eff} = M_{eff}/L^2$ denotes the effective reservoir area density. The formula reflects that fluctuations increase at smaller length scales $\xi$.

Next, we estimate the length scale over which such fluctuations give rise to macroscopic coherence. When as a result of a fluctuation the density becomes high enough that quantum degeneracy is reached (see also the representations of Figs. 3a,b), the wavepacket size $\xi$ in this regime is approximately given by $\xi = \left(\lambda_{th}/\sqrt{4\pi}\right)e^{n\lambda_{th}^2/2}$ [39]. This formula can be used to estimate the density $n_{deg}$ above which quasi long-range ($\geq \lambda_{th}$) coherence emerges, yielding a phase-space density of $n_{deg}\lambda_{th}^2 \simeq \ln(4\pi) \simeq 2.53$. Therefore, the emergence of a quantum degenerate microcloud is expected for $\Delta n \geq n_{deg} - \bar{n}$, since the local density can then become sufficiently large that a wavepacket with extended coherence over a distance $\xi$ can form (see Figs. 3a,b).

The surface density required to establish coherence over some length scale $\xi$ (with $\xi \geq \lambda_{th}$ given by $n(\xi) = \frac{2}{\lambda_{th}^2}\ln\left(\xi\sqrt{4\pi}/\lambda_{th}\right)$. For an average photon density $\bar{n}$, coherence over this length scale can be achieved when $\Delta n = \sqrt{2\rho_{eff}}/\xi \geq n(\xi) - \bar{n}$. From this, we find that wavepackets with extended coherence of size ξ, triggered by thermal fluctuations, can form for $\xi \lesssim \xi_{max}$, as determined by the relation $\xi_{max} = \sqrt{2\rho_{eff}} \Big/ \left[\frac{2}{\lambda_{th}^2} \ln\left(\frac{\lambda_{th}}{\xi_{max}\sqrt{4\pi}}\right) - \bar{n})\right]$. Across the full 2D quantum gas, we expect a transiently emerging pattern of local quantum degenerate wavepackets (see Fig. 4). and the total source emission is not expected to possess macroscopic coherence due to both spatial and temporal averaging.

Quite similarly, for average densities above the one required to establish long range coherence over the whole system size in the absence of fluctuations (i.e. for $\bar{n} > n(\xi = L)$), we expect that grand canonical fluctuations can cause defects in an otherwise macroscopically coherent photon cloud. These fluctuations lead to local undershoots of the density and correspondingly reduced coherence, in particular when the density drops below the value at which quantum degeneracy is reached. In the course of this, also topological defects may emerge, comparable to effects observed, e.g. during quenches through phase transitions [40]. Here, however, such defect dynamics are not induced by an external quench but are continuously driven by intrinsic grand-canonical fluctuations.

It is important to emphasize that the effects described here differ from the phase fluctuations known for low-dimensional systems described in earlier works [39]. In the present case, they arise from the grand canonical nature of the system, specifically due to the possibility of local interconversion between photons and material electronic excitations. Naturally (see also above), the precondition for these effects to become relevant is that no spatial averaging of the photon density due to transverse motion in the resonator plane suppresses these effects, i.e. we require a 'hydrodynamic' regime. With $\overrightarrow{p_{op}} = \frac{\hbar}{i}\vec{\nabla}$ acting on the above ansatz for the wavefunction for a localized photon wavepacket, we arrive at a transverse velocity $v = p/m \simeq \hbar/(m\xi) \cdot e^{-x/\xi}$ along position x. Hence, reabsorption of photons typically needs to occur at a timescale $t_{reabs} \lesssim \xi/v \simeq \xi^2 m/\hbar$. For say $\xi$=5μm, we arrive at a required reabsorption time $t_{reabs} \lesssim 1$ ps, as for the example of rhodamine dye solution of 1mmol/l concentration is feasible for cutoff wavelengths $\lambda_{cutoff} \lesssim 560$ nm, i.e. closer to the spectral maximum of absorption than used in most current dye microcavity photon condensation experiments [5, 14-16, 37,38].

## IV.) Conclusions

In conclusion, we have introduced a free energy model to describe macroscopic coherence of photons in material-filled optical microcavities coupled to a reservoir of material electronic excitations. A single mode treatment of condensate photons, as applicable e.g. to present harmonically trapped dye microcavities, for the noninteracting gas yields the usual Mexican-hat shaped dependence of the free energy when written versus the real and imaginary parts of the complex wave function for a small reservoir, corresponding to the canonical statistical regime. For a large reservoir, statistical number fluctuations increase as the grand canonical regime is approached. Here, for increased reservoir sizes we predict a regime with a simple bowl-shaped dependence of the free energy centred at the origin, for which no spontaneous symmetry breaking upon condensation is expected.

Furthermore, we considered a microcavity photon gas with spatially distributed interconversion between photons and material electronic excitations, representing a generalisation of the above model to also include the spatial degrees of freedom. In this case, patterns of transiently emerging islands with extended coherence are predicted, each of them with an independent phase, and triggered by thermal fluctuations.

For the future, refined experimental studies of the coherence properties of photon condensates to simultaneously determine first and second-order coherence properties deeply in the grand canonical regime are highly desirable. Beyond dye microcavity condensates, we note that photon condensates coupled to a gaseous reservoir, such as proposed noble gas mixture UV microcavity condensates [41], where the density of the optically excitable gas atoms can potentially be higher than that of dye molecules, are expected to operate more deeply in the grand canonical regime than current dye-based systems. For practical applications in the visible and near infrared spectral regimes, recently developed semiconductor microcavity based photon condensates, with setups that closely resemble VCSEL-laser sources, are particularly promising platforms [18, 19].

Prospects of this work include the exploration of novel radiation sources whose coherence properties are broadly tuneable between laser-like and lamp-like behavior, in terms of first-order (both transverse and longitudinal) as well as second-order coherence. Potential technical applications include speckle-free imaging and optical coherence tomography.

We thank Jan Klaers for helpful discussions. This work was financially supported by the Deutsche Forschungsgemeinschaft within SFB/TR 185 (277625399) the Cluster of Excellence ML4Q (390534769), and the EU (ERC, TopoGrand, 101040409). K.P. acknowledges financial support by the Walter Benjamin programme of the DFG (CONDENS, 567107833)

*present address: CNR Nano, Istituto Nanoscienze, Piazza San Silvestro 12, 56127 Pisa, Italy

**Literature:**

[1] M. H. Anderson, J. R. Enscher, M. R. Matthews, C. E. Wieman, and E. A. Cornell, Science **269**, 198 (1995).

[2] K. B. Davis et al., Phys. Rev. Lett. **75**, 3969 (1995).

[3] J. Kasprzak et al., Nature **443**, 409 (2006).

[4] R. Balili, V. Hartwell, D. Snoke, K. Pfeiffer, and K. West, Science **316**, 1007 (2007).

[5] J. Klaers, J. Schmitt, F. Vewinger, and M. Weitz, Nature **468**, 545 (2010).

[6] A. Einstein, Sitz.ber. Preuss. Akad. Wiss. **1**, 3 (1925).

[7] N. Bogoliubov, Journal of Physics (USSR) **11**, 23 (1947).

[8] O. Penrose and L. Onsager, Phys. Rev. **104**, 576 (1956).

[9] P. W. Anderson, Phys. Rev. **112**, 1900 (1958).

[10] L. Pitaevskii and S. Stringari, *Bose-Einstein condensation and superfluidity* (Oxford University Press, Oxford, 2016).

[11] J. Bloch, I. Carusotto, and M. Wouters, Nature Rev. Phys. **4**, 470 (2022).

[12] M. R Andrews, C. G. Townsend, H.-J. Miesner, D. S. Durfee, D. M. Kurn, and W. Ketterle, Science **275**, 637 (1997).

[13] J. Schmitt, T. Damm, D. Dung, C. Wahl. F. Vewinger, J. Klaers, and M. Weitz, Phys. Rev. Lett. **116**, 033604 (2016).

[14] J. Marelic and R. A. Nyman, Phys. Rev. A **91**, 033813 (2015).

[15] S. Greveling, K. L. Perrier, and D. van Oosten, Phys. Rev. A **98**, 013810 (2018).

[16] M. Vretenar, C. Toebes, and J. Klaers, Nature Commun. **12**, 5749 (2021).

[17] R. Weill, A. Bekker, B. Levit, and B. Fischer, Nat. Commun. **10**, 747 (2019).

[18] R. C. Schofield et al., Nature Photonics **18**, 1083 (2024).

[19] M. Pieczarka et al., Nature Photonics **18**, 1090 (2024).

[20] I. Fujiwara, D. Ter Haar, and H. Wergeland, J. Stat. Phys. **2**, 329 (1970).

[21] R. M. Ziff, G. E. Uhlenbeck, M. Kac, Phys. Rep. **32**, 169 (1997)

[22] M. Holthaus, E. Kalinowski, and K. Kirsten, Annals of Physics **270**, 198 (1998).

[23] J. Klaers, J. Schmitt, T. Damm, F. Vewinger, and M. Weitz, Phys. Rev. Lett. **108**, 160403 (2012).

[24] D. N. Sob'yanin, Phys. Rev. E **85**, 061120 (2012).

[25] M. Kruk et al., Rep. Prog. Phys. **88**, 106401 (2025)

[26] A. Crisanti, A. Sarracino, and M. Zannetti, Phys. Rev A **113**, 033307 (2026).

[27] J. Schmitt, T. Damm, D. Dung, F. Vewinger, J. Klaers, and M. Weitz, Phys. Rev. Lett. **112**, 030401 (2014).

[28] B. T. Walker, J. D. Rodrigues, H. D. Dhar, R. P. Oulton, F. Mintert, and R. A. Nyman, Nature Commun. **11**, 1390 (2020).

[29] P. Kirton and J. Keeling, Phys. Rev. A **91**, 033826 (2015).

[30] W. Verstaelen and M.Wouters, Phys. Rev. A **100**, 013804 (2019).

[31] V. Yukalov, Laser Phys. Lett. **34**, 113001 (2024).

[32] F. Öztürk et al., Science **372**, 6537 (2021).

[33] F. Öztürk, F. Vewinger, M. Weitz, and J. Schmitt, Phys. Rev. Lett. **130**, 033602 (2023).

[34] See also the results obtained for the phase jump rate and the second-order coherence times respectively in Ref. 13.

[35] A. Abouelela et al, Phys. Rev. Lett **136**, 053402 (2025).

[36] J. Schmitt, J. Phys. B **51**, 173001 (2018).

[37] E. Busley et al., Science **375**, 1403 (2022).

[38] L. Kleebank et al., Sci. Adv. **12**, eaee2942 (2026).

[39] Z. Hadzibabic and J. Dalibard, Riv. Nuovo Cimento **34**, 389 (2011).

[40] L. E. Sadler, J. M. Higbie, S. L. Leslie, M. Vengalattore, and D. M. Stamper-Kurn, Nature **443**, 312 (2006).

[41] E. Boltersdorf, T. vom Hövel, J.A. Morín Nenoff, F. Vewinger, and M. Weitz, Phys. Rev. A **112**, 012811 (2025).

**Figures:**

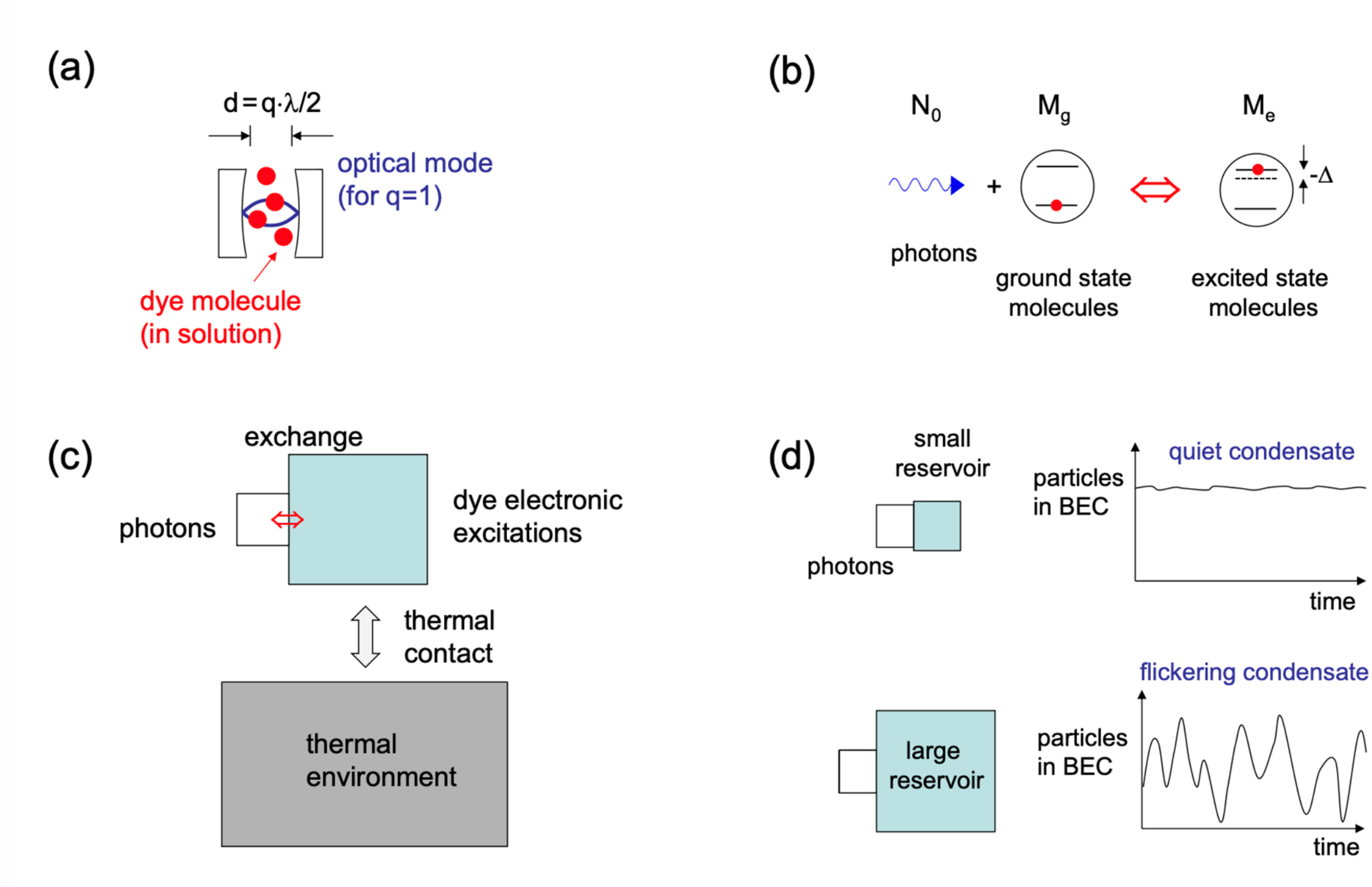

Fig. 1 (a): Scheme of an optical microresonator with harmonic trapping of a two-dimensional photon gas, for the case of a distance between resonator mirrors of half the optical wavelength of cavity photons (q = 1). The interior of the resonator is filled with a dye solution, and the photon gas thermalizes through absorption and emission processes on the dye molecules. In this process, the longitudinal modal quantum number q is frozen out and only the two transversal degrees of freedom are varied. (b) Photons can be converted into electronically excited state dye molecules and vice versa. (c) Representation of used calculational model. The interconversion of photons and dye excitations can be described as an effective particle exchange between condensate photons and dye electronic excitations. Further, the dye is in thermal contact with the solvent, which acts as a thermal reservoir. (d) The magnitude of statistical number fluctuations depends on the size of the dye reservoir. Grand canonical conditions are fulfilled for a large relative reservoir size (bottom), leading to strong photon bunching.

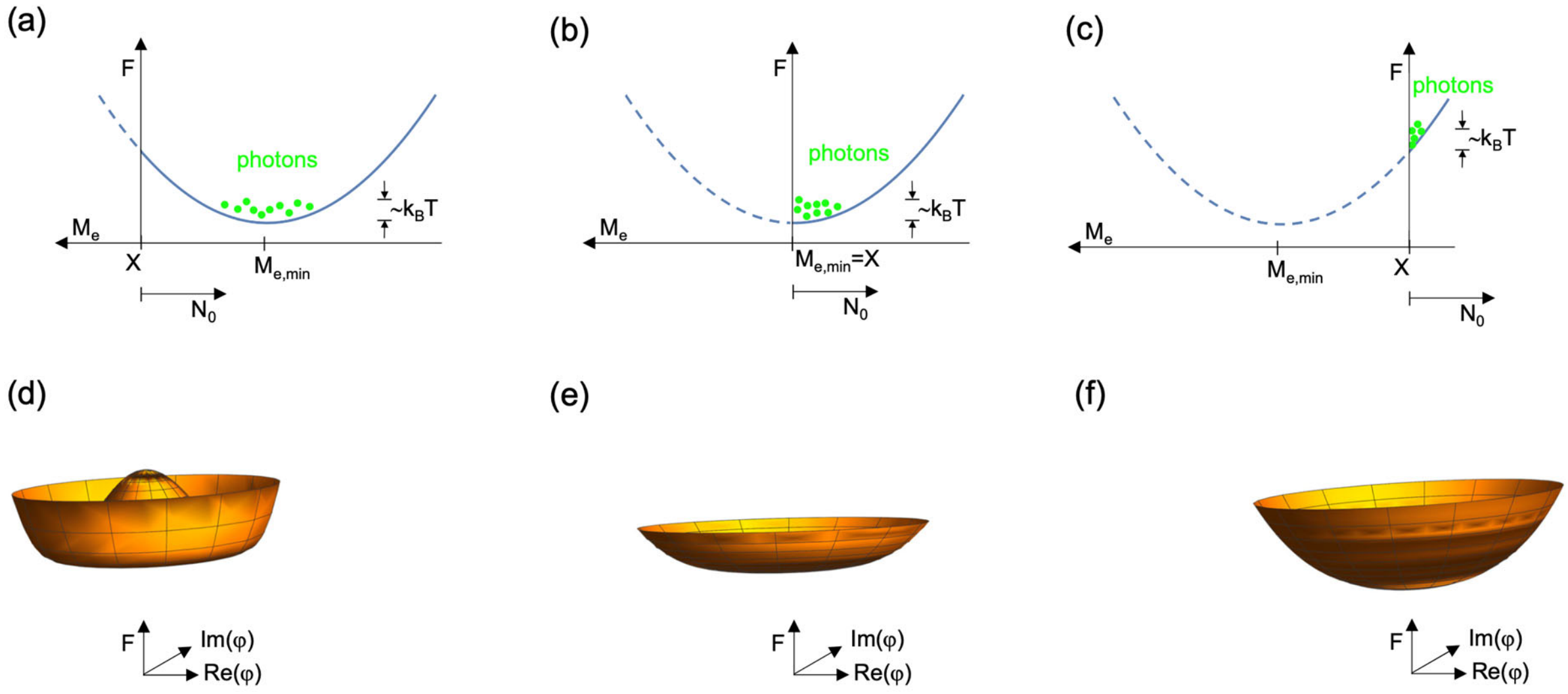

Fig. 2: Variation of the free energy $F$ of photon condensate coupled to dye reservoir versus both the upper electronic state molecular population $M_e$ (axis to the left) and condensate photon number $N_0 = X - M_e$ (axis to the right) for (a) $X > M_{e,min}$, (b) $M_{e,min} = X$, and (c) $X < M_{e,min}$. The excitation number $X$ gives the total number of excitations, and unphysical parts of the free energy curve are shown as a dashed line. For (a) and (c) it has been assumed that $F(X) - F(M_{e,min}) >> k_BT$, such that the diagrams illustrate the free energy variations in (a) the canonical respectively (c) far grand canonical statistical regimes, while (b) gives the transition point between the two regimes. (d, e, f) Free energy versus the real and imaginary parts of the condensate instantaneous wave function φ for (d) canonical, (e) the transition point, and (f) the far grand canonical cases. In the latter case, a macroscopic ground state population in the absence of spontaneous symmetry breaking is expected.

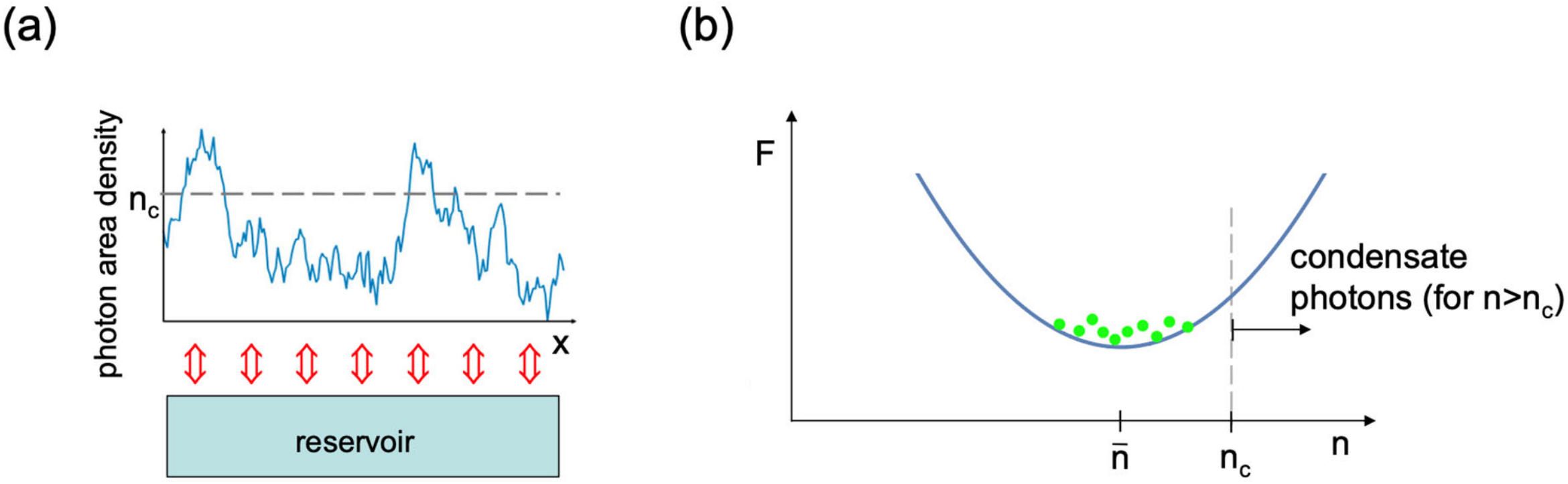

Fig. 3: (a) Cut though the spatial profile of a statistically varying photon area density in an optical microresonator for a spatially distributed coupling to the dye reservoir. The dashed horizontal line shows the area density $n_{deg} \simeq \ln(4\pi)/\lambda_{th}^{2}$ at which the onset of long range coherence ($\xi \geq \lambda_{th}$) is reached. (b) Variation of the free energy $F$ in an area $\xi^2$ of spatial uniform conversion between photons and dye electronic excitations on the local area density $n$ (see text). When the area density exceeds the onset value for the quantum degenerate regime, long-range coherence emerges.

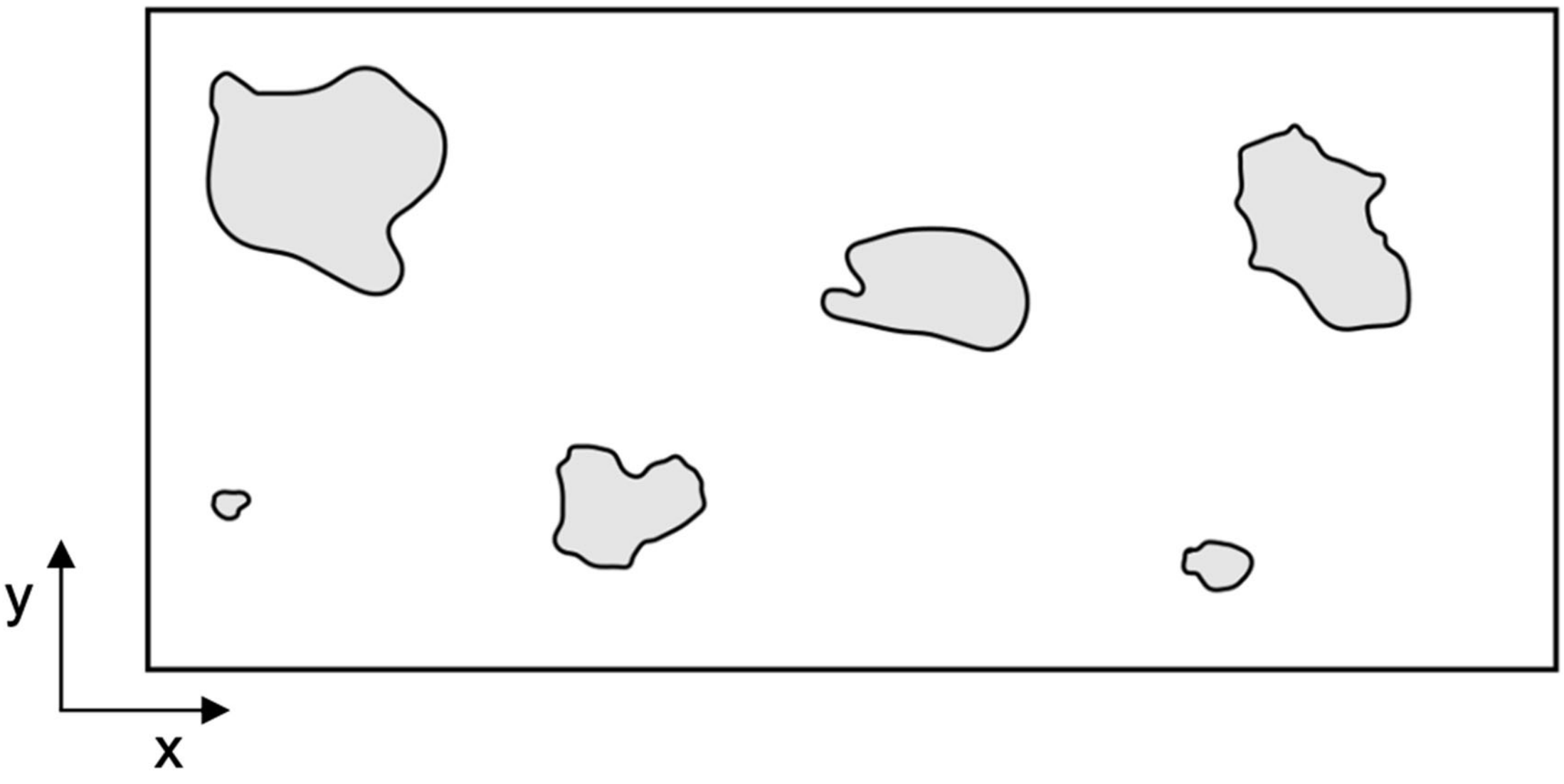

Fig. 4: Schematic representation of instantaneous emission pattern of a two-dimensional photon gas, with a pattern of "microcondensates" emerging by grand canonical statistical fluctuations.